# Chapter 27
# Risk-Based Prognostics and Health Management

John W. Sheppard
Montana State University
Bozeman, Montana

## 27.1. Introduction

As engineering fields mature, new technologies are emerging that are beginning to serve as the foundation of many societal improvements. For example, modern medical diagnostic equipment provides valuable information that gives medical professionals a better understanding of a patient's needs and ultimately improves quality of life [1]. Improvements to vehicle designs make transportation in cars or aircraft safer and more environmentally friendly [2]. Military equipment continues to be developed that better supports and protects personnel in the field [3]. Manufacturing practices and robotic equipment improve work safety conditions and reduce a product's price point, making amenities available to a wider range of consumers [4].

One approach to maximizing system availability is to incorporate some means of health assessment into the system itself. Doing so is often referred to as "integrated system health management" (ISHM) or "prognostics and health management" (PHM), which has been applied successfully to many complex systems [5]. By integrating health assessment into the very functioning of a system, more information can be obtained that provides a better understanding of the system as a whole, thus allowing system owners to become proactive in how they deal with system degradation. ISHM and PHM promise to focus on system conditions, thus supporting initiatives in what has become known as condition-based maintenance (CBM). This, in turn, enables maintenance events to be initiated based on specific system conditions rather than waiting until a failure occurs [6].

One of the key ingredients of ISHM/PHM is diagnostics, which corresponds to the process of determining the health state of the system based on sets of observations (or tests). Such tests are designed specifically to track system behavior and determine whether or not a failure has occurred. In many cases it is impossible to identify a single fault that explains the observations with certainty. Instead, candidate sets of faults are often indicated, and when using applicable models, probabilities or confidence values are associated with the faults to provide additional information. One historic approach to using test observations for diagnosis is to apply a decision tree – sometimes referred to as a fault tree[1] [7]. In a diagnostic fault tree, a path through the tree corresponds to a sequence of test results that ends in identifying a fault, replaceable unit, or ambiguity group.

---

[1] For most of this chapter, we will use an alternative version of a fault tree that characterizes how faults propagate to effects and associated hazards in a system.

A companion process to diagnostics is prognostics. With prognostics, the goal is not only to determine the current health state of a system but also to predict the health state at some time in the future [8]. This prediction is key to the "P" part of "PHM." Similar to diagnostics, exact identification of this future health state is rare, and a probabilistic approach is desirable quantify the level of uncertainty associated with the health state predictions more reliably. When moving beyond detecting incipient failures (another common task in PHM), a temporal component must be introduced into the models. A variety of model types have been proposed for this, including the Kalman filter [9][10], the particle filter [11][12], the dynamic Bayesian network [13][14], and recently the continuous time Bayesian network (CTBN) [15][16].

In real-world, complex system support, PHM introduces significant challenges, not only with how to predict faults but also with how support organizations measure the effectiveness of maintenance processes that act upon such predictions. To address this issue, we propose coupling the prognostic process explicitly, rather than implicitly, with a risk-based framework. Specifically, we begin by observing that faults, by their very nature, produce undesirable effects in the system, which we refer to as hazards. With the ability to predict the health state of a system at some point in time, it is possible to anticipate faults and the risks associated with resulting hazards. This information can then be used to assess the severity or criticality of the risk and potentially mitigate that risk through preventative maintenance. Effectiveness measurement then shifts to be in terms of hazards averted. Alternative measures based on overall availability can also be used for such assessment.

### 27.2. Risk-Based PHM

While much work in PHM has focused more on the health management aspects, a large part of PHM involves prognostics. Such approaches often use a prognostic model to predict when a system fault is likely to occur, based on an understanding of the current health of that system. Identifying the most critical faults is achieved by domain experts synthesizing the prognostic results, who then use these faults and existing reliability data to identify likely hazards [17]. Finally, based on the determined likely effects, the domain expert makes a decision about the best course of action, given the predicted behavior.

Although this method has proven itself to be relatively effective, there are several shortcomings of this approach. The first is the heavy reliance on data synthesis by human experts. This can be a problem in that the process requires highly skilled domain experts to understand the details of the system whenever information needs to be collected and analyzed. Relying on the human expert, however, is problematic given the goals for real-time health assessment and prediction, and the steady increase in system complexity. Thus, minimizing the need for human involvement in PHM is critical to the success and integration of PHM into real-time system support.

Another concern regarding current PHM approaches is that the process often involves a pipeline of largely disjoint models. In particular, a diagnostic model is used to assess observations and test results to determine the health of the system. This health information is then provided to a prognostic model, which then predicts the future occurrence of potential faults. An entirely separate model is then used to identify which effects will be most impacted by those faults. Finally, yet another model, set of equations, or domain knowledge is used to evaluate the potential impact of emerging hazards on the system based on the likelihood of the different effects and other behaviors. Each step in the process depends on separate technologies, making implementation and integration difficult. Furthermore, information passed between technologies tends to be transcribed manually, unless a standard data exchange format is observed. This disjointed approach is inefficient and has the potential for information loss due to model incompatibilities. A better solution would seamlessly work with all tasks necessary to support PHM in responding to emerging hazards and their associated risks.

One approach to performing "risk-informed decision making" that was developed by the United States Nuclear Regulatory Commission is embodied in a tool called Systems Analysis Programs for Hands-on Integrated Reliability Evaluations (SAPHIRE) [18]. The SAPHIRE system has been designed to support large-scale analysis of fault trees and event trees to determine how faults propagate to effects. SAPHIRE uses a process for generating minimal cut sets for the fault trees, thereby reducing their complexity to the most critical set of faults. The associated probabilistic analyses focus on deriving importance measures such as Fussell-Vesely importance and risk achievement, which are used to assess probabilistic properties of risk in a system [19]. Ultimately, SAPHIRE's intended use is static, intended to be used to assess the risk of a design; it is not typically used for online risk assessment or PHM.

In this chapter, we introduce risk-based PHM (rPHM), a cohesive, online methodology and framework that supports a variety of PHM tasks using a single type of model. This framework supports both diagnostics and prognostics and incorporates effects or hazards using the same model semantics. By building hazards into the model itself, predictions can be made not only about likely faults but also about the effects that may occur as a result of those faults. The integration of faults, observations, and hazards is handled algorithmically and avoids the need for real-time human synthesis. Furthermore, the framework supports the definition of user-specified functions that place value on various system states, thus providing a means for quantifying system quality while still using the semantics of the underlying prognostic model. The framework also allows decisions (also referred to as "scenarios") to be modeled, which means that multiple system options can be compared to one another to facilitate risk mitigation. Finally, by incorporating performance functions into the model, a multi-objective approach to optimization can be used to assess possible courses of action and to facilitate more effective risk-informed decision-making.

This chapter provides an overview of techniques we have developed to define the rPHM framework. Section 27.3 provides background information on the models and data structures used in the rPHM framework. Sections 27.4 and 27.5 explain the procedures necessary to build the rPHM models using readily available reliability information. Section 27.6 goes into detail concerning how the rPHM model is intended to be used, drawing on our work in performance functions and continuous-time decision networks. The main points of the chapter are summarized in Section 27.7.

## 27.3. Background

In this section, we review the models and data structures used throughout this chapter.

### 27.3.1. Continuous Time Bayesian Networks

Many systems change state as a function of time, and the continuous time Bayesian network (CTBN) is a framework that is capable of modeling such systems [20]. The CTBN represents each variable in a system as vertices in a directed graph. These vertices are connected using directed edges such that the behavior of each vertex depends directly on the behavior
of its parents in the graph.

To describe the behavior of a variable, the CTBN defines a conditional Markov process for the vertex. This consists of an initial distribution defined over the states of the variable, as well as a conditional intensity matrix (CIM) that describes how the variable changes its state over time. More formally, a CIM $\mathbf{Q}_X$ for some variable (vertex) $X$ specifies a set of transition intensities $q_{i,j}(\mathbf{Pa}(X))$, where $q_{i,j}$ is the entry in the $i^{\text{th}}$ row, $j^{\text{th}}$ column of $\mathbf{Q}_X$, which change as a function of the states assigned to the parents of $X$ ($\mathbf{Pa}(X)$). These entries specify the main parameter to an exponential distribution, $f(t) = q_{i,j} \exp(-q_{i,j} t)$. When specifying the CIM, we also note that the entry on the diagonal is given as $q_- = \sum_j q_{i,j}$.

The resulting CTBN defines a distribution over the states of each variable at any given time $t \in [0, \infty)$. Approximate inference algorithms have been developed that are capable of answering queries about state probabilities or transition times tractably. These algorithms work with continuous time evidence, which allows the states of variables to be set for specified time intervals (e.g., during maintenance). In this chapter, we use the CTBN as the basis for supporting PHM since it provides a model by which we can predict the probability of faults and hazards based on the evidence provided by the tests.

### 27.3.2. Notation

We begin our formal treatment of the CTBN model by providing an overview of notation we use through the remainder of this chapter. Capital letters $F_i$ and $T_j$ denote faults and tests respectively,, where the subscript indicates the specific test or fault in the system. A bold $\mathbf{F_i}$ is a vector of faults detected by some test $T_i$, and $\mathbf{F_i}[j]$ indicates the $j^{\text{th}}$ fault detected by $T_i$. We use a bold $\mathbf{T_j}$ to denote a vector of tests that detect a fault $F_j$. This corresponds to what is known as the "diagnostic signature" for that fault. The symbols $\lambda_i$ and $\mu_i$ specify the failure and repair rates respectively for fault $F_i$. Similarly, $ND_{i,j}$ and $FA_{i,j}$ also describe non-detect and false-alarm probabilities, but they do so for the specific test $(T_i, \mathbf{F_i}[j])$ pair. This allows us to capture the notion that a test may fail to detect a particular fault while remaining operational. A complete list of notation used in this chapter is provided in Table 3.1.

| Symbol | Interpretation |
|---|---|
| $F_i$ | The $i^{\text{th}}$ fault in the network |
| $T_i$ | The $i^{\text{th}}$ test in the network |
| $\mathbf{F_i}$ | The vector of faults monitored by $T_i$ (i.e., the parent set of $T_i$) |
| $\mathbf{T_i}$ | The vector of tests that will detect the failure of $F_i$ |
| $\mathbf{D}$ | A D-matrix |
| $d_{i,j}$ | The $\langle F_i, T_j \rangle$ entry of 0 or 1 in a D-matrix |
| $\lambda_i$ | The failure rate for fault $F_i$ |
| $\mu_i$ | The repair rate for fault $F_i$ |
| $ND_i$ | The non-detect probability for test $T_i$ |
| $FA_i$ | The false alarm probability for test $T_i$ |
| $ND_{i,j}$ | The non-detect probability for test-fault relationship $T_i$ and $\mathbf{F_i}[j]$ |
| $FA_{i,j}$ | The false alarm probability for test-fault relationship $T_i$ and $\mathbf{F_i}[j]$ |
| $\mathbf{Q}_X$ | The intensity matrix for variable $X$ |
| $\mathbf{Q}_{X\|Y}$ | The conditional intensity matrix for variable $X$, conditioned on variable $Y$ |
| $\mathbf{Pa}(X)$ | The parents of variable $X$ |

*Table 3.1 Notation used in this chapter.*

*27.3.3. D-Matrices*

In order to determine the health state of a system, whether for purposes of fault diagnosis or prognosis, we need certain information about the unit under test (UUT). This information is collected via tests performed on the UUT, whether through the use of built-in test, on-board health monitoring, off-board automatic testing, or manual testing. The outcome of these tests then indicate the presence or absence of faults in the system [7]. In a complex system, the way tests and faults interact with one another can make diagnosis and maintenance difficult since a single fault may be monitored by multiple tests, and a single test may monitor multiple faults. To manage this complexity, we represent the relationships between the faults and tests explicitly using a representation that corresponds to a graph adjacency matrix, known as a D-matrix. With this matrix, the columns correspond to tests and the rows correspond to the faults or failures observed by these tests. The D-matrix has been adopted by several modeling tools used to perform fault diagnosis, including the aerospace community [21]. In fact, the concept of a "fault dictionary," which is used in digital testing, can be regarded as a special case of a D-matrix [22][23].

Constructing a D-matrix involves building diagnostic signatures for each fault within the UUT. Given a set of faults **F** and tests **T**, we define the matrix D as follows:

$$d_{i,j} = \begin{cases} 1 & \text{if } T_i \text{ detects } F_j \\ 0 & \text{otherwise} \end{cases}.$$

For some fault $F_i$, the diagnostic signature corresponds to $\mathbf{F}_i = [d_{i,1}, ..., d_{i,|\mathbf{T}|}]$ for that fault [21][24]. This diagnostic signature is determined for each $F_i \in \mathbf{F}$, and each signature $\mathbf{F}_i$ becomes the $i^{\text{th}}$ row of the D-matrix.

The relationships between tests and faults described by a D-matrix provide information needed to diagnose the UUT, but probabilistic information is not included in this structure [24]. Similarly, the basic D-matrix does not provide information about fault-to-fault or test-to-test relationships; although, test-to-test relationships can be derived through a process known as "logical closure" [7]. For these reasons, while the D-matrix can be used to construct a network structure (Section 27.4.1), more information is required to parameterize a CTBN in order to perform diagnosis and prognosis (Section 27.4.2).

*27.3.4. Fault Trees*

Fault tree analysis (FTA) is a powerful method for evaluating system design in a reliability and safety-oriented context. In several critical domains, fault trees provide a graphical representation of how failures of one or more system components can lead to higher-order system failures. A fault tree is a directed acyclic graph (DAG) consisting of a set of events and a set of logic gates. Interior vertices of the graph represent effects, while the leaf vertices are faults or base-level failures. Leaf states propagate upward through the tree towards the root, capturing how faults contribute to possible system effects. Each interior vertex receives input from its children, which are then combined through a logic gate such as an AND gate or an OR gate.

Some fault tree representations allow for redundancies in the system that can cause the same component to appear in multiple parts of the tree, thus violating the strict tree structure of the underlying graph. When these redundant components are removed from the tree through a pruning process, the resulting fault tree becomes easier to store, update, and process since the pruned tree remains functionally equivalent to the original tree.

Note that many of these redundancies can be removed by finding the minimal cut sets of the fault tree. Since minimal cut set analysis is focused on identifying the cut sets affecting the top-level effect of the tree, we found that this type of analysis is not suitable for our purposes. This is because we are also interested in the occurrence of the intermediate effects as well [25].

A related diagnostic model is the Fault Isolation Manual (FIM), which often plays a significant role in maintaining large systems [7]. FIMs are derived from decision trees that are also typically referred to as fault trees[2]. The internal vertices of the FIM correspond to tests, that are represented in the D-matrix, and the leaves correspond to faults, also represented in the D-matrix. To use the FIM, the test specified at the root of the tree is performed, and the result of that test indicates which branch to follow, leading to the next vertex in the tree. This process continues until reaching a leaf corresponding to a diagnosed fault or ambiguity group. Thus, following the decision tree structure in a FIM leads to isolating faults or ambiguity groups [24].

## 27.4. Deriving CTBNs from D-Matrices

Given that our work focuses on performing fault prognosis as a foundation, we begin by considering how information derived from a diagnostic model—namely the D-matrix—can be used as the first step in creating our rPHM model.

### 27.4.1. D-Matrix Network Structure

The non-zero entries in a D-matrix represent dependence relationships between tests and diagnoses (e.g., faults). Similarly, the directed edges in a CTBN indicate influence between variables. This observation can be used to derive the network structure of a CTBN from a D-matrix. Let **D** be an $m \times n$ D-matrix, and let $\mathcal{G}$ be a graph structure for a CTBN (initially containing no vertices). First, $m$ fault vertices are added to $\mathcal{G}$, each one representing a row in **D**. Then, $n$ test vertices are added, representing the columns in **D**. Finally, for each entry in the D-matrix where $d_{i,j} = 1$, an edge is inserted from $F_i$ to $T_j$, where $F_i$ is the fault vertex corresponding to the $i$th row in **D**, and $T_j$ is the test vertex associated with the $j$th column in **D**. Each of these edges assert that the behavior of a test depends on the state of a fault at some point in time.

The resulting network $\mathcal{G}$ has $m + n$ vertices and $m_d$ edges, where $m_d = \sum_{i=1}^{m} d_{i,j}$. These directed edges are inserted from a fault vertex to a test vertex, with no edges inserted from tests to faults, between two test vertices, or between two fault vertices. The resulting graph $\mathcal{G}$ is therefore bipartite, with one layer consisting of the $m$ faults and the other consisting of the $n$ tests.

Typically the variables contain relatively few states, meaning that the initial probability distributions and intensity matrices for each vertex are small. As a result, the total number of intensity matrices defined over the set of all vertices is the primary factor driving model complexity. None of the fault vertex will have parents, thus only a single intensity matrix is required to parameterize each fault. The test vertices, however, may have any number of fault vertices in their parent sets. As a result, a CIM for a test vertex will have $\left|\mathbf{Q}_{T_j}\right| = \prod_{i=1}^{m} d_{i,j} \cdot |F_i|$, matrices, where $|F_i|$ is the number of states for $F_i$, and $d_{i,j}$ equal to one indicates the presence of a dependence. In the worst case, a test may monitor all faults in the system, resulting in an upper bound $O(cm)$, where $c$ serves as a bound on the number of states for each fault. In fault diagnosis, it is common for $c = 2$; however, this could increase should multiple states, perhaps representing levels of degradation, be included.

---

[2]Note that FIM-based fault trees are different from fault trees developed through FTA.

*27.4.2. D-Matrix Parameterization*

The way one variable influences another, as well as default failure transition behavior, is captured in the CTBN by parameterizing each vertex. One way of doing such parameterization is by using reliability information; however, these parameters could be updated as data is collected during the operation and maintenance of the system. For the purpose of this chapter, we assume binary states for both fault and test vertices; however, the same ideas could be applied to variables with more than two states if supporting data is available.

1) *Parameterizing Fault Vertices:* The fault vertices in the network can be parameterized independent of the test vertices since they have no parents in the network. This leads to the requirement to specify only a single unconditional intensity matrix for each fault. Recall that the intensity matrix defines exponential transition distributions over the states. In other words, the CIM specifies the probability distributions associated with transitioning from a non-failed state to a failed state and vice-versa. Failure and repair rates are often available via either historical data or domain knowledge. Formally, a failure rate $\lambda$ indicates the rate at which a failure will occur given that no failure currently exists. Note that the reliability analysis often captures information about mean time between failure (MTBF), and that with the exponential distribution assumption, is defined as $MTBF = 1/\lambda$.

The repair rate $\mu$ indicates the rate at which a fault will transition back to having no failure, and depending on the associated repair policy is often set to zero. In that case, the failure state will become an absorbing state. In general, when parameterizing a fault's intensity matrix, the failure rate $\lambda$ is assigned to the entry corresponding to the transition from non-failure to failure, and the repair rate $\mu$ is assigned to the reverse transition (i.e., failure to non-failure). The the diagonal values are set to be the negative sum of the remaining entries in the row (i.e., $-\lambda$ and $-\mu$ respectively). Let $F_i$ be a fault vertex. Then the intensity matrix for $F_i$ is

$$\mathbf{Q}_{F_i} = \begin{array}{c} \\ f_i^0 \\ f_i^1 \end{array} \begin{array}{c} f_i^0 \quad f_i^1 \\ \begin{pmatrix} -\lambda_i & \lambda_i \\ \mu_i & -\mu_i \end{pmatrix} \end{array}$$

where $\lambda_i$ and $\mu_i$ are the failure and repair rates for $F_i$.

Note that in some cases, transition rates follow nonexponential distributions. In reliability modeling, the Weibull and lognormal distributions are two distributions that are commonly used to describe time-to-failure [26]. While a CTBN is defined based on exponential distributions, it is possible to approximate non-exponential distribution using "phase-type distributions" [27]. Furthermore, work has been done to obtain approximations of non-exponential parametric distributions automatically, thereby allowing for more complex failure distributions to be represented. The corresponding phase-type distribution is then learned and embedded directly into the existing matrix $\mathbf{Q}_{F_i}$.

2) *Parameterizing Test Vertices:* Parameterizing test vertices is more complex than parameterizing the fault vertices, in that test vertices have $n \geq 1$ parents in the graph, meaning that the number of intensity matrices required to specify a test CIM grows exponentially with respect to $n$.

The intensity matrices for a test vertex's CIM can be derived using false alarm (FA) and non-detect (ND) likelihoods for each test. FA and ND values can be estimated based on sensitivity and calibration information on the test. The relationship between FA and ND rates and the probability distribution for a test can then be modeled as line failures, described in detail by Perreault *et al.* [15]. The resulting probability distribution is given as

$$P(T_i = 0|\mathbf{F}_i) = (ND_i)P(p_i = 1|\mathbf{F}_i) + (1 - FA_i)P(p_i = 0|\mathbf{F}_i),$$

where $T_i$ is a test in the network, $F_i$ is the set of parent faults for $T_i$, and $ND_i$ and $FA_i$ are non-detect and false alarm probabilities respectively, defined for $T_i$.

Note that the distribution incorporates two probabilities $P(p_i = 1|\mathbf{F}_i)$ and $P(p_i = 0|\mathbf{F}_i)$, which can be calculated as follows:

$$P(p_i = 1|\mathbf{F}_i) = \prod_{\{f_{i,j} \in \mathbf{F}_i | f_{i,j}=0\}} (1 - FA_{i,j}) \cdot \prod_{\{f_{i,j} \in \mathbf{F}_i | f_{i,j}=1\}} (ND_{i,j})$$

$$P(p_i = 0|\mathbf{F}_i) = (1 - P(p_i = 1|\mathbf{F}_i))$$

Here, $ND_{i,j}$ and $FA_{i,j}$ correspond to the non-detect and false alarm probabilities for test $T_i$ with respect to the $j^{\text{th}}$ fault in the parent set of $T_i$. The entire probability distribution $(p_i|\mathbf{F}_i)$ can be thought of as the probability that the true state of faults $\mathbf{F}_i$ are detected correctly (or incorrectly) by test $T_i$.

Given the previous equation, the number of non-detect and false alarm probabilities to be modeled is equal to the number of edges in the CTBN network, plus the total number of tests: $O(|E| + |V|)$. Although this number is quite manageable from a computational standpoint, it may be that no data exists to describe the needed probabilities for the individual fault-test relationships. Instead, specific fault-test values $ND_{i,j}$ and $FA_{i,j}$ may be assumed to be zero, and any false alarm or non-detect likelihoods can be summarized by $ND_i$ and $FA_i$ for test $T_i$.

With this assumption, the number of user-specified parameters is reduced drastically. The only non-detect and false alarm probabiliteis that need to be specified are for the test vertices: $O(|V|)$. If we assume a test fails in the presence of any failure of the monitored components, then the test's output will follow a deterministic OR gate. Then the intensity matrix corresponding to no faults in the system becomes

$$\mathbf{Q}_{\{T_i | \wedge_{F \in \mathbf{F}_i} F=0\}} \begin{array}{c} \\ t_i^0 \\ t_i^1 \end{array} \begin{pmatrix} t_i^0 & t_i^1 \\ -(1-FA_i)^{-1} & (1-FA_i)^{-1} \\ (FA_i)^{-1} & -(FA_i)^{-1} \end{pmatrix}.$$

and the intensity matrix for the cases with faults becomes

$$\mathbf{Q}_{\{T_i | \vee_{F \in \mathbf{F}_i} F=1\}} \begin{array}{c} \\ t_i^0 \\ t_i^1 \end{array} \begin{pmatrix} t_i^0 & t_i^1 \\ -(ND_i)^{-1} & (ND_i)^{-1} \\ (1-ND_i)^{-1} & -(1-ND_i)^{-1} \end{pmatrix}.$$

Then each matrix in CIM $\mathbf{Q}_{T_i}$ can be parameterized based on the high-level false alarm and non-detect probabilities. Again, this information can be learned based on historical maintenance data.

## 27.5. Deriving CTBNs from Fault Trees

The approach to prognostics espoused in this chapter is to model hazards that might arise in a system and capture the associated risks. The approach taken is based on the fault tree analysis discussed in Section 27.3.4. In this section, we present a method for deriving CTBNs from fault trees, resulting in a new predictive model with faults and hazard-based effects. We then show how to relate this information to the D-matrix and explain how to merge the two networks.

### 27.5.1. Fault Tree Network Structure

The structure of a hazard/effect-based CTBN can be derived directly using the fault tree developed through FTA as follows. First the set of vertices **V** is obtained by extracting the variables directly from the faults and hazards in the fault tree. Recall that the hazard/effect nodes correspond to logic gates. Note that either a fault or hazard may occur in multiple locations of the fault tree, but the corresponding vertex in the CTBN will occur only once. Thus, the resulting CTBN structure may not be defined strictly as a tree.

Next, directed edges are inserted between the vertices. Once again, these can be obtained directly from the structure of the fault tree. First recall that faults only occur as leaves in the tree. The corresponding vertices in the CTBN, as with the D-matrix-based CTBN, therefore have no parents. Next we consider the hazards in the tree whose states are determined as a logical combination of their inputs. We represent this dependence on the inputs by adding an edge from the vertices corresponding to each input $u_i$ to the vertex corresponding to the effect. The resulting CTBN has the same underlying structure as the fault-tree after pruning. As mentioned, collapsing multiple occurrences of faults into a single vertex will result in a graph that is no longer a tree. The pruning process may yield substantially fewer edge insertions, and given that the space complexity of a vertex is exponential with respect to the number of parents, pruning is a critical preprocessing step.

### 27.5.2. Fault Tree Parameterization

In system reliability, the prior probability that each component starts in a failing state is often known, and in this chapter, we assume all variables start in a non-failed state with probability 1.0. In other words, this forms the prior probability for the fault vertices. As before, we assume the variables are binary, consisting of a failing state and a non-failing state. Then the intensity matrix for each vertex $X$ requires $2^{(|\mathbf{Pa}(X)|+1)}$ rates, which can make identifying parameters for even relatively small models a difficult task. Just as in Section 27.4, the goal is to reduce the number of rates required to parameterize a vertex by taking advantage of available reliability information and by exploiting behavioral information obtained from the underlying fault tree. The fault vertices in the fault tree-based network are identical to those obtained from the D-matrix; therefore, the same parameterization process from Section 27.4.2 can be employed.

*1) Parameterizing AND Effects:* Let $\Phi_X$ be the discrete function corresponding to the gate from which variable $X$ was derived. Then we expect $\lim_{t \to \infty} P\left(X(t) = \Phi_x(\mathbf{Pa}(X))\right) = 1.0$. This behavior is guaranteed for each CIM $Q_{X|\mathbf{Pa}(X)}$ by forcing a non-zero transition rate *to* the state produced by $\Phi_X(\mathbf{Pa}(X))$, and a zero rate *from* the state produced by $\Phi_X(\mathbf{Pa}(X))$. In other words, this creates an absorbing state, at which point the variable will remain in this state, unless a change in its parent states occurs.

For a vertex created corresponding to an AND gate, $\Phi_X$ becomes the logical AND function, where $\Phi_X$ will produce a value of 1 if and only if all inputs are 1. To accomplish this, the CIM is parameterized for a vertex $X$ where all parents are 1 as follows:

$$\mathbf{Q}_{X|\mathbf{Pa}(X)} = \begin{matrix} \\ x^0 \\ x^1 \end{matrix} \begin{pmatrix} x^0 & x^1 \\ -\lambda_X & \lambda_X \\ 0 & 0 \end{pmatrix}.$$

when $\Phi_X(\mathbf{Pa}(X)) = \mathbf{1}$ (i.e., all parent states are ones).

Considering the cases where not all parents of the vertex are 1, the function $\Phi_X$ produces a value of 0, so we wish to describe the time it takes to transition back to a state of 0 for the vertex in the CTBN. For this, the CIM is parameterized as:

$$\mathbf{Q}_{X|\mathbf{Pa}(X)} = \begin{matrix} \\ x^0 \\ x^1 \end{matrix} \begin{pmatrix} x^0 & x^1 \\ 0 & 0 \\ \mu_{X|\mathbf{Pa}(X)} & -\mu_{X|\mathbf{Pa}(X)} \end{pmatrix}.$$

when $\Phi_X(\mathbf{Pa}(X)) = \overline{\mathbf{1}}$ (i.e., one or more of the parents is in state 0).

Note that we need to specify a rate for transitioning back to 0 for each of the instances where not all parents are in state 1 since each of these cases has a separate intensity matrix. Although this level of detail may be necessary in some cases, it may be possible to simplify the parameterization process by assuming the rate is the same, regardless of the parent set. This means that $X$ transitions back to 0 at the same rate as long as $\Phi_X$ evaluates to 0. Given this, a variable $X$ derived from an AND vertex in a fault tree can be specified using only two parameters, where $\lambda_X$ defines the time it takes to transition to 1 in the event that all parents are in state 1, and $\mu_X$ is used in all other cases to indicate when $X$ will transition back to 0.

*2) Parameterizing OR Effects:* When a variable $X$ is created from an OR gate, $\Phi_X$ is defined as the logical OR function. Again, it is guaranteed that this CTBN vertex eventually reaches state 0 by parameterizing the intensity matrix such that state 0 is an absorbing state:

$$\mathbf{Q}_{X|\mathbf{Pa}(X)} = \begin{matrix} \\ x^0 \\ x^1 \end{matrix} \begin{pmatrix} x^0 & x^1 \\ 0 & 0 \\ \mu_X & -\mu_X \end{pmatrix}.$$

when $\Phi_X(\mathbf{Pa}(X)) = \mathbf{0}$ (i.e., all of the parent vertices are in state zero).

Next, consider the case where $\Phi_X$ evaluates to 1. Since $\Phi_X$ is an OR gate, this occurs whenever at least one parent is in state 1. In this case, vertex $X$ is parameterized such that the matrices in the CIM eventually transition to state 1, ensuring that $X$ conforms to the correct output of the OR gate:

$$\mathbf{Q}_{X|\mathbf{Pa}(X)} = \begin{matrix} \\ x^0 \\ x^1 \end{matrix} \begin{pmatrix} x^0 & x^1 \\ -\lambda_{X|\mathbf{Pa}(X)} & \lambda_{X|\mathbf{Pa}(X)} \\ 0 & 0 \end{pmatrix}.$$

when $\Phi_X(\mathbf{Pa}(X)) = \overline{\mathbf{0}}$ (i.e., one or more of the parents is in state one).

Once again, we need to specify a rate parameter for every possible state instantiation of the parents where at least one parent is in state 1. This means that $2^{|\mathbf{Pa}(X)|}$ parameters would required for vertex $X$. As in the case of the AND vertex, this number can be reduced by making a simplifying assumption. Given the causal interpretation of the model, the concept of disjunctive interaction, also referred to as the Noisy-OR model, may be employed [28]. This allows us to specify the rates $\lambda_{X_i}$ for only the cases where a single parent is in state 1. This reduces the number of required parameters to be linear in the number of parents. The remaining parameters for the CIMs with multiple parents in state 1 are then accounted for by disjunctive interaction.

*27.5.3. Merging the D-Matrix and Fault Tree Models*

Now that we have a way to construct a CTBN that captures diagnostic and prognostic relationships and a way to create a separate CTBN that models effect/hazard propagation, the next step is to combine them into a single model. The key observation that makes this merge process possible is that both the D-matrix-derived CTBN and a fault tree-derived CTBN describe different aspects of the same single system $S$; the set of faults is the same for both models.

Let $CTBN_D$ be a CTBN constructed from the D-matrix for system $S$, and let $CTBN_F$ be a CTBN derived from the fault tree for the same system $S$. As before, let **T** be the set of test vertices in $CTBN_D$, **H** be the set of hazard vertices in $CTBN_F$, and **F** the set of faults contained in both CTBNs. A new model $CTBN_S$ can be constructed for system $S$ by combining both networks to obtain a single CTBN with vertices **T**, **H**, and **F**. The vertices in the sets **T** and **H** will have the same parents and parameters as in the original networks. The only difference is that the vertices in the set **F** in the combined CTBN have more children than they did before being merged, but this does not affect their parameterization. Furthermore, the fault vertices in $CTBN_D$ and $CTBN_F$ are parameterized using the same method.

## 27.6. Usage and Decision Making

A CTBN model can use observations collected from tests to perform diagnostics and prognostics. More specifically, we can regard the test results (which indicate whether or not the tests passed or failed, as well as when the test was run) as evidence and then apply that evidence to the test vertices. We can then query the model to determine probability distributions over the states of faults and effects at any point in time (both future and past). These probability distributions can then be used to identify the most likely effects and how likely they are with an eye towards mitigating the risks associated with the hazards.

Falling under the umbrella of what is called "risk-informed decision making," the rPHM approach described here is intended to use the derived probability information about emerging risks to help operators and maintainers do their jobs more effectively. Specifically, our approach is to incorporate performance directly into the model and use standard query techniques to quantify the overall evolution of the state of the system, relative to health, risk, and potential impact on mission effectiveness. An elegant way of achieving this while still maintaining the benefits of the CTBN's factored representation is to make use of what we refer to as "performance functions," which extend the standard model by allowing more sophisticated, user-defined queries [29]. Essentially, a value is assigned to the time associated with entering or remaining in a system state, and this value can be aggregated over multiple vertices and throughout the entire time of interest. The remainder of this section describes how the quantities from these user-defined functions can be used to help decide how best to respond to that evolving system state.

### 27.6.1. Handling System Scenarios

Here, we will assume that we have constructed a CTBN consisting of a D-matrix and a fault tree. For this discussion, we will make use of a running example. Specifically, suppose we have modeled a vehicle that contains the subsystems shown in Table 6.2. Notice that this table also includes MTBF and MTTR, from which the various parameters for $\lambda$ and $\mu$ can be derived respectively. Suppose also that we seek to mitigate the following hazards

1. Loss of Power
2. Loss of Electrical
3. Loss of Chassis
4. Loss of Engine
5. Loss of Power Train
6. Loss of Vehicle
7. Loss of Crew

| Subsystems | Subsystem Name | MTBF | MTTR | Repair Cost |
|---|---|---|---|---|
| AI | Air System | 900 | 5 | 250 |
| AL | Alternator | 1300 | 15 | 900 |
| AX | Axles | 1800 | 12 | 2000 |
| BR | Brakes | 1200 | 5 | 950 |
| CO | Cooling System | 2100 | 6 | 200 |
| EL | Electronics | 1700 | 20 | 600 |
| FU | Fuel System | 800 | 16 | 1200 |
| IG | Ignition System | 600 | 13 | 300 |
| PR | Compression System | 1100 | 35 | 9000 |
| PW1 | Power Source 1 | 400 | 3 | 250 |
| PW2 | Power Source 2 | 200 | 3 | 250 |
| SU | Suspension System | 1500 | 8 | 850 |
| TR | Transmission | 2500 | 30 | 6500 |
| WT | Wheels & Tires | 700 | 3 | 700 |

*Table 6.2. Summary of Subsystems in Vehicle Model*

In some cases, it may be useful to construct multiple versions of the same model. We refer to these model versions as "scenarios", examples of which could correspond to maintenance actions or different possible system operational modes. For instance, one possible situation might be to perform preventative maintenance on a vehicle at time $t = 500$ where the maintainer would replace the wheels and tires before this subsystem fails. Addressing this would require our model to include a subsystem (represented as a vertex) corresponding to "wheels and tires," which we denote as WT. This situation results in two possible scenarios: 1) do not perform a preventative action, and 2) replace the WT component. In the replacement scenario, downtime for the system will result in order to perform the maintenance, and there will be a cost associated with the maintenance process itself. But the goal of such proactive maintenance would be to reduce the likelihood that the subsysstem fails in the near term, which might in turn prevent a more drastic Loss of Vehicle event. Indeed, it might even prevent a Loss of Crew event in the extreme case.

Another example of a scenario would be where we need to change the operational mode of the vehicle. For example, suppose the cooling system has a high probability of failing in the near future. Then rather than continuing to operate the vehicle as normal, a more conservative approach could be taken that reduces speed and conserves power consumption. This would then reduce the likelihood of failure of the Cooling System (or adjacent subsystem), but it might have other costs associated with the action. For example, by reducing power, the time to get on target would likely be prolonged.

There are several options for modeling scenarios. As suggested above, the first is to duplicate the base model and make the necessary changes to the network structure or parameters to account for the differences in the various scenarios. The result is a set of networks (i.e., models) that vary in some way based on the specifics of the scenarios. Queries can then be performed over the set networks simultaneously, and the results can be compared. Although this works well in theory, scenarios often differ by very little, meaning that the corresponding networks will have several redundancies, thus potentially increasing the computational complexity associated with processing such similar models. Furthermore, it is possible that scenarios depend upon other choices, which can lead to an exponential explosion in the number of models required. This would especially be a concern when there are many scenarios, when the models are very large, or when there are several complex interactions among the scenarios. Furthermore, in the event that a change must be made to the model, this change would need to be propagated through all of the copied networks, raising the risk of the resulting models being inconsistent.

An alternative approach that addresses some of these issues is to introduce a new to the model. We refer to this new vertex as a "decision vertex." This vertex acts identically to a standard CTBN vertex, but it is used in a specific way to handle the various scenarios while preserving model commonality across the scenarios. Each state of a decision vertex corresponds a different decision or scenario that may be "selected," and we assume the state of the decision vertex is known at all times. Multiple decision vertices can be included in the same model. Each of the decision vertices is defined with no parents, but any vertices that are affected by the scenario are added to the child set of the decision vertex. Then the child vertex's parameters (i.e., CIMs) are adjusted to account for the different states of the scenario.

More specifically, the approach is to insert a new parent node on the fault, subsystem, or effect/hazard of interest, where that parent node reflects the conditions for that subsystem in that scenario. For example, suppose we are considering the Cooling System (CO) situation described above and we define a parent decision vertex labeled "Operation." For the purposes of this example, assume the vehicle has two operational modes: standard and conservative. We recognize that this violates the constraint imposed previously that each fault has no parents, and what this means is that the components of the vehicle no longer behave independently of external factors. Specifically, the external factors correspond to the decisions made to determine the appropriate scenario (i.e., setting the operational mode).

The model must now be reparameterized to account for this change to the network's structure. Note that subsystem CO originally had a single CIM with only one intensity matrix $\mathbf{Q}_{CO}$. This assumed standard operation of the vehicle. The new CIM now needs two matrices to be defined, one for each state of its new parent.

$$\mathbf{Q}'_{CO|\text{Operation}} = \{\mathbf{Q}'_{CO|\text{Standard}}, \mathbf{Q}'_{CO|\text{Conservative}}\}.$$

Since the original model assumed standard operation, the transition behavior in that case is already known and $\mathbf{Q}'_{CO|\text{Standard}} = \mathbf{Q}_{CO}$. The remaining intensity matrix $\mathbf{Q}'_{CO|\text{Conservative}}$ then needs to be defined to describe the new cooling system behavior given a conservative operational mode. For this scenario, the failure rate would likely decrease with the conservative driving style due to the reduced wear-and-tear on the vehicle. Note, however, that the repair rate would remain the same. Since the state of the decision vertex is known at all times, no inference will need to be performed to obtain this state. Thus, there is no need to parameterize this vertex[3]. The only requirement to maintain mathematical consistency is that all states of the decision vertex are controllable at any given time, meaning that the initial distribution must have non-zero entries, and a transition may not occur with infinite intensity. For example, we could use a uniform initial distribution over the states and zero transition rates.

To use the model extended to include decision vertices, inference can be run over the network or each possible decision, and the results compared. For our example, that means fixing the Operation vertex to be in the standard state and running inference, and then repeating the process after setting the Operation vertex to the conservative state. To aid in comparing the results, performance functions can be defined [29]. Let $\pi_{Op}$ be a performance function that assigns a positive value for every hour spent in the operational or partially operational state and no value when in a non-operational state. For example, we could define the following.

$$\pi_{Op} = \begin{cases} 10t & \text{if } LOV = \text{false} \land Operation = \text{standard} \\ 5t & \text{if } LOV = \text{false} \land Operation = \text{conservative} \\ 0 & \text{otherwise} \end{cases}$$

where $LOV$ is shorthand for Loss of Vehicle

Inference can then be run by considering the two options available, and the value of each option can be quantified by querying over function $\pi_{Op}$. While running the vehicle in a conservative state is less valuable than standard operation, the decrease in the failure rate for the cooling system could be substantial enough to prevent a total Loss of Vehicle event. In fact, it could mitigate a significant risk. In particular, where under standard operation there would be a lower cost of operation (due to the shorter time required), the mission might have a high probability of failure. On the other hand, under the conservative model, there would be an increase in cost (due to the increase in time required), but it would have a higher probability of success. It might therefore make more sense to take the partial value that is available rather than risk losing the vehicle entirely and thereby failing the mission.

Performance functions can be very complex; they can be defined to use more sophisticated equations or incorporate other risks that are of interest, such as the Loss of Crew hazard. Multiple performance functions can also be defined, each attempting to address a different objective or goal, thus setting up a multi-objective optimization problem.. With these functions built directly into the rPHM framework, however, it is possible not only to support diagnostics and prognostics, but also to compare scenarios using metrics that are relevant to the system's application. After inferring performance values for each possible scenario, choosing the best scenario reduces to the task of optimizing over those values [30]. In the multi-objective case, methods exist to obtain a Pareto front of non-dominated decisions. This subset of Pareto-optimal options can then be presented to a domain expert who can decide what is best for the given context [31][32].

---

[3] It is possible to have adaptive scenarios, in which case such parameterization and inference would be required. Note that the CTBN model already supports this, but it is beyond the scope of the discussion here.

*27.6.2. Performance-Based Logistics*

In addition to supporting decisions for runtime operations, the rPHM models discussed in this chapter can be used during system design as well. This is especially useful when working under the objectives of performance based logistics (PBL), which is a contracting strategy that aims to improve operational effectiveness in large organizations like the US Department of Defense. A more detailed treatment of what follows can be found in [33]. Unlike other approaches that contract for resources, PBL contracts for performance according to a variety of prespecified criteria [34]. These criteria are related directly to system performance and factors such as reliability, maintainability, and supportability [35]. In short, PBL contracts require contractors and suppliers to meet life-cycle performance criteria, thus requiring an objective method for evaluating performance against those criteria.

By building a rPHM CTBN during the the design and development process, accurate predictions of system behavior can be made. Furthermore, performance functions can be constructed and included directly in the model that relate to the objectives laid out in the PBL contract. Often, these objectives incorporate the impact of failure to achieve mission objectives and are well suited to incorporating risk-based modeling. Finally, different design alternatives can be added to the model using the decision vertices discussed in the previous section. This allows design objectives (including in the multi-objective sense) to be evaluated with respect to the contract requirements prior to actual implementation, saving time and money for the contractor, and improving the overall solution for the end user.

To demonstrate this idea, we continue to use our hypothetical vehicle model. During the design phase, there may be a choice between two different axle systems.

Table 6.2 shows that the AX component has an MTBF of 1800 hours, an MTTR of 12 hours, and a repair cost of 2000. Suppose there is a second axle system AX′ to be considered that has an MTBF of 2100 hours, an MTTR of 20 hours, and a repair cost of 2500. This data reflects that the new axle design is more complex (due to the higher repair cost), but that this increase in complexity results in higher reliability and therefore a longer mean time between failure. Unfortunately, in addition to the higher repair cost, the additional design complexity also increases the mean repair time. Then the question that the contractor is left with is which design alternative to choose to best meet the contract requirements.

The correct choice between these axle design alternatives is not immediately obvious and depends on a number of factors, including how changes to reliability impact the associated risks and what performance criteria are most important for the contract. To approach this from the rPHM perspective, the vehicle CTBN model can be modified to incorporate the various design options. Similar to the way we inserted a parent vertex Operation into our model, suppose we define a parent vertex AxleChoice, which we denote $D_{AX}$. We make this the parent of the AX vertex in the network. To parameterize the AX vertex, an intensity matrix is derived once for each design alternative and paired with the corresponding state in the $D_{AX}$ decision vertex. Once again, the parameters in the decision vertex are largely unimportant since the state will be assigned deterministically. With this model, inference can be run once for each design alternative, allowing the performance values to be inferred for each objective. Alternatives whose performances do not meet the contract requirements can be eliminated, and a design choice can be obtained from the remaining valid solutions employing multi-objective decision making methods [32].

In addition to parameter changes, decision vertices can be used to support structure changes to the associated network as well. For example, there may be two design alternatives concerning the design of the power subsystem, where one option includes a redundant power source for fault tolerance while the other alternative proceeds from a single power source. For example, suppose instead of a single power source, we have two power sources, denoted PW1 and PW2 respectively. Here, we place a decision node $D_{PW}$ as the parent of the Loss of Power event. Assuming PW1 and PW2 are both binary, then the CIM for Loss of Power originally consisted of four intensity matrices corresponding to the redundant power source scenario. These four intensity matrices are mapped to the first state of $D_{PW}$, and four new intensity matrices are mapped to the second state that covers the single power source scenario. With these four new matrices, Loss of Power is parameterized according to PW1 alone as if PW2 did not exist. When $D_{PW}$ is in the first state (redundant design), Loss of Power is dependent on both PW1 and PW2. When $D_{PW}$ is in the second state (no redundancy), Loss of Power is dependent on only PW1, and the edge between PW2 and Loss of Power is effectively severed.

These models provide a mathematically sound and inexpensive method for evaluating design choices with respect to contract requirements prior to implementation. In addition, once the system is deployed, the model can be used for diagnostics, prognostics, and online risk mitigation. This is achieved by fixing the decision vertices representing design choices to the alternatives that were chosen during operation. Going forward, these models can then be used to identify fault and effect likelihoods, as well as being used to evaluate specified performance functions. Additional decision nodes can be inserted to model additional operational decisions, which can serve as mitigation strategies that could minimize performance loss.

## 27.7. Summary

In this chapter, we introduced the concept of risk-based prognostics and health management, with a specific focus on developing a CTBN framework for implementing this concept. To reduce the barrier to entry when building the CTBN models, we introduced a number of procedures that naturally generate the structure and parameters of the model using commonly available diagnostic and reliability information. By explaining how both structure and parameters can be obtained using this data, we are able to eliminate the need to apply expensive learning algorithms or manual modeling processes, while at the same time providing a modeling framework that could make use of machine learning methods, if desired. We then explained how to extend the modeling framework to support capturing different scenarios and demonstrated how to apply them with a hypothetical vehicle model.

In addition to detailing processes for building CTBN-based prognostic models, we discussed how the resulting models can be used to perform diagnostics and prognostics by applying evidence corresponding to test results. Furthermore, we presented methods for representing and reasoning over operational decisions to support making design decisions and to incorporate strategies for mitigating risk. Specifically, we argued for incorporating decision vertices and performance functions into the CTBN, which enable representing various scenarios and utilities, can be optimized to evaluate alternatives and make risk-informed decisions during runtime. Finally, we showed that these models were not only useful for making operational decisions, but could also be applied to design time decisions as well. By providing a framework for optimizing various design decisions, these models can support tasks like PBL while also continuing to serve as prognostic models capable of describing runtime behavior after deployment.